\documentclass[%
 aip,
rsi,%
 amsmath,amssymb,
 reprint,%
]{revtex4-1}

\usepackage{graphicx}
\usepackage{dcolumn}
\usepackage{bm}
\usepackage[colorinlistoftodos]{todonotes}

\begin{document}


\title{An Undergraduate Test of Gravitational Time Dilation}

\author{Brian Patterson}
\email{brian.patterson@usafa.edu}
\affiliation{Department of Physics, US Air Force Academy, Colorado 80840}

\author{Mario Serna}
\affiliation{Department of Physics, US Air Force Academy, Colorado 80840}

\author{Jerry F. Sell}
\affiliation{Department of Physics, US Air Force Academy, Colorado 80840}

\author{M. Alina Gearba}
\affiliation{Department of Physics, US Air Force Academy, Colorado 80840}

\author{Robert Olesen}
\affiliation{Department of Physics, US Air Force Academy, Colorado 80840}

\author{Patrick O'Shea}
\affiliation{Department of Physics, US Air Force Academy, Colorado 80840}

\author{Jonathan Schiller}
\affiliation{Department of Physics, US Air Force Academy, Colorado 80840}

\author{David Emanuel}
\affiliation{Department of Physics, US Air Force Academy, Colorado 80840}

\author{M. Shane Burns}
\email{sburns@coloradocollege.edu}
\affiliation{Department of Physics, Colorado College, Colorado Springs, Colorado 80903}

\author{Michael D. Leveille}
\affiliation{Department of Physics, Colorado College, Colorado Springs, Colorado 80903}

\author{Armand R. Dominguez}
\affiliation{Department of Physics, Colorado College, Colorado Springs, Colorado 80903}

\author{Brian B. Gebhard}
\affiliation{Department of Physics, Colorado College, Colorado Springs, Colorado 80903}

\author{Samuel E. Huestis}
\affiliation{Department of Physics, Colorado College, Colorado Springs, Colorado 80903}

\author{Jeffrey Steele}
\affiliation{Department of Physics, Colorado College, Colorado Springs, Colorado 80903}

\date{\today}

\begin{abstract}
Students at Colorado College and cadets at the US Air Force Academy have conducted an experimental test of gravitational time dilation.  This relativistic effect, highlighted in the movie \textit{Interstellar}, causes clocks to tick more slowly near massive objects. A measurement of gravitational time dilation was made by comparing signals generated by a GPS frequency standard, which is based on sea-level time, to a cesium-beam frequency standard located at three different elevations above sea level. The effect is small but observable; for the highest elevation studied (4302 m on the summit of Pikes Peak), a local clock ticks only 41 ns/day faster than a sea-level clock. Our results are consistent with the predictions of general relativity. We also discuss implications of gravitational time dilation on GPS operations.
\end{abstract}

\maketitle

%

According to Einstein's general theory of relativity, a clock runs more slowly if it is close to a large, gravitating object. This principle was highlighted in the movie \textit{Interstellar}, in which the main character spends several hours on a planet orbiting  a massive black hole, and returns to find that his young daughter has become an elderly woman. In the movie, one hour on the planet equates to seven years on Earth. The same effect can be measured on a much smaller scale by changing locations within Earth's gravitational field. Using atomic clocks, we have compared the ticking of time at Colorado College (elevation 1845 m above sea level), at the United States Air Force Academy (elevation 2165 m), and on the summit of Pikes Peak (elevation 4302 m). Our results confirm that clocks do indeed tick more quickly at higher altitudes, in accordance with Einstein's theory. This work was carried out as a collaboration of physics majors and faculty at Colorado College and the US Air Force Academy. The experimental technique is straightforward and our results can be reproduced by other students and teachers. Indeed, a similar endeavor has been reported by amateur clock enthusiasts.\cite{vanbaak2007}  

\section*{Background}
Albert Einstein first postulated the notion of gravitational time dilation in 1907 using the principle of equivalence.\cite{einstein1907} He showed that the local time, $\tau$, as measured by a clock at a specified location in a homogeneous gravitational field, can be written approximately as
\begin{equation}
   \tau = \tau_0 \left( 1+\frac{\Phi}{c^2} \right).
   \label{genrel}
\end{equation}
In Eq.~\ref{genrel}, $\Phi$ is the gravitational potential at the specified location, $\tau_0$ is the time measured by a clock at an arbitrarily chosen origin of coordinates, at which $\Phi$ is defined to be zero, and $c$ is the speed of light. Thus two clocks, at two different locations, will keep the same time only if they experience the same gravitational potential. For our measurements we chose an origin at sea level, corresponding to the time-keeping convention of GPS.  For a homogeneous gravitational field, the gravitational potential is simply $\Phi = g h$, where $g$ is the local acceleration due to gravity and $h$ is the altitude with respect to sea level. We can therefore write the fractional time difference between two clocks, one at sea level and one at altitude $h$, as
\begin{equation}
   \label{approxeqn1}
   \frac{\tau-\tau_0}{\tau_0} = \frac{g \, h}{c^2} .
   \end{equation}
In applying Eq.~\ref{approxeqn1}, the value of $g$ must be approximately constant over the distance $h$, which is valid for the three locations considered here. At Colorado College ($h = 1845$ m), the fractional time dilation is $2 \times 10^{-13}$ or $17$ ns per day. That is, each day clocks at Colorado College tick an extra $17$ ns compared to clocks at sea level. Similarly, the fractional time dilations at the Air Force Academy and on the summit of Pikes Peak are $21$ ns/day and $41$ ns/day, respectively. If larger altitudes are examined, such that $g$ cannot be considered constant, Eq.~\ref{approxeqn1} must be integrated over the altitude range of interest using the altitude dependence of $g$.

Clocks at different altitudes travel at slightly different speeds due to their differing distances from Earth’'s rotation axis. It is worth considering, therefore, whether special relativistic effects are significant. According to the well-known time dilation formula\cite{knight2016} in special relativity, a clock moving at speed $v$ with respect to an observer will measure a time
\begin{equation}
   \tau = \tau_0 \sqrt{1-v^2/c^2}
   \label{specrel}
\end{equation}
compared to a clock at relative rest, which measures time $\tau_0$. For speeds much smaller than the speed of light, this gives a fractional time dilation of
\begin{equation}
   \frac{\tau-\tau_0}{\tau_0} = \frac{v^2}{2c^2} .
   \label{approxeqn2}
\end{equation}
For the altitudes considered in this paper, the special relativistic effect is small ($< 1$ ns/day) and will be ignored. At satellite altitudes, however, where orbital speeds are on the order of a few km/s, this effect is comparable in magnitude to gravitational time dilation, as will be discussed later. It is worth noting that Eqs.~\ref{approxeqn1} and \ref{approxeqn2} can also be derived from Einstein’'s full theory of general relativity using the Schwarzschild metric.\cite{ashby2003,burns2017}

\section*{Experimental Setup}

\begin{figure}
\includegraphics[width=0.4\textwidth]{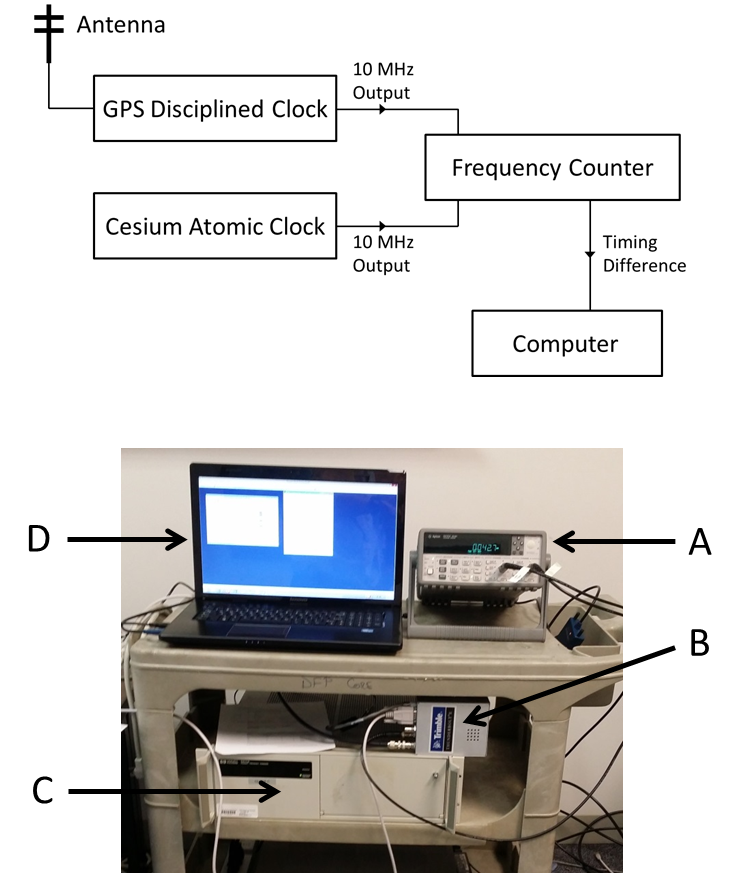}
\caption{\label{expsetup} Experimental Setup. A frequency counter (A) measures the timing difference between output signals from a GPS disciplined clock (B) and a cesium atomic clock (C). The timing data are recorded using a laptop computer (D). }
\end{figure}

The upper panel in Fig.~\ref{expsetup} shows a schematic of the experimental setup. The local time is measured using a HP 5071A cesium frequency standard (atomic clock), which outputs a nominal 10-MHz timing signal. A Trimble Thunderbolt GPS disciplined clock, locked to multiple GPS satellites, provides a standard GPS timing signal (also 10 MHz) referenced to sea-level. Because the atomic clock is located several thousand meters above sea level, there is a small but measurable difference in the frequencies of the two timing signals. During the course of the experiment, a phase shift continually accumulates between the two signals, which we measure using an Agilent 53000 Series frequency counter operated in time-interval mode. The frequency counter measures the time difference between upward zero-crossings of the two timing signals. The data are recorded every 1-2 seconds using a simple computer program.  A date-time stamp is also recorded for each data point. The data are acquired continuously for 4-5 days in a typical data run. Multiple runs were made at Colorado College (CC), the Air Force Academy (USAFA), and on the summit of Pikes Peak (PP). The lower panel in Fig.~\ref{expsetup} shows the actual experimental hardware. The cesium clock is about the size of a large stereo receiver, so the entire setup is quite portable. 

\section*{Results}

\begin{figure}
\includegraphics[width=0.4\textwidth]{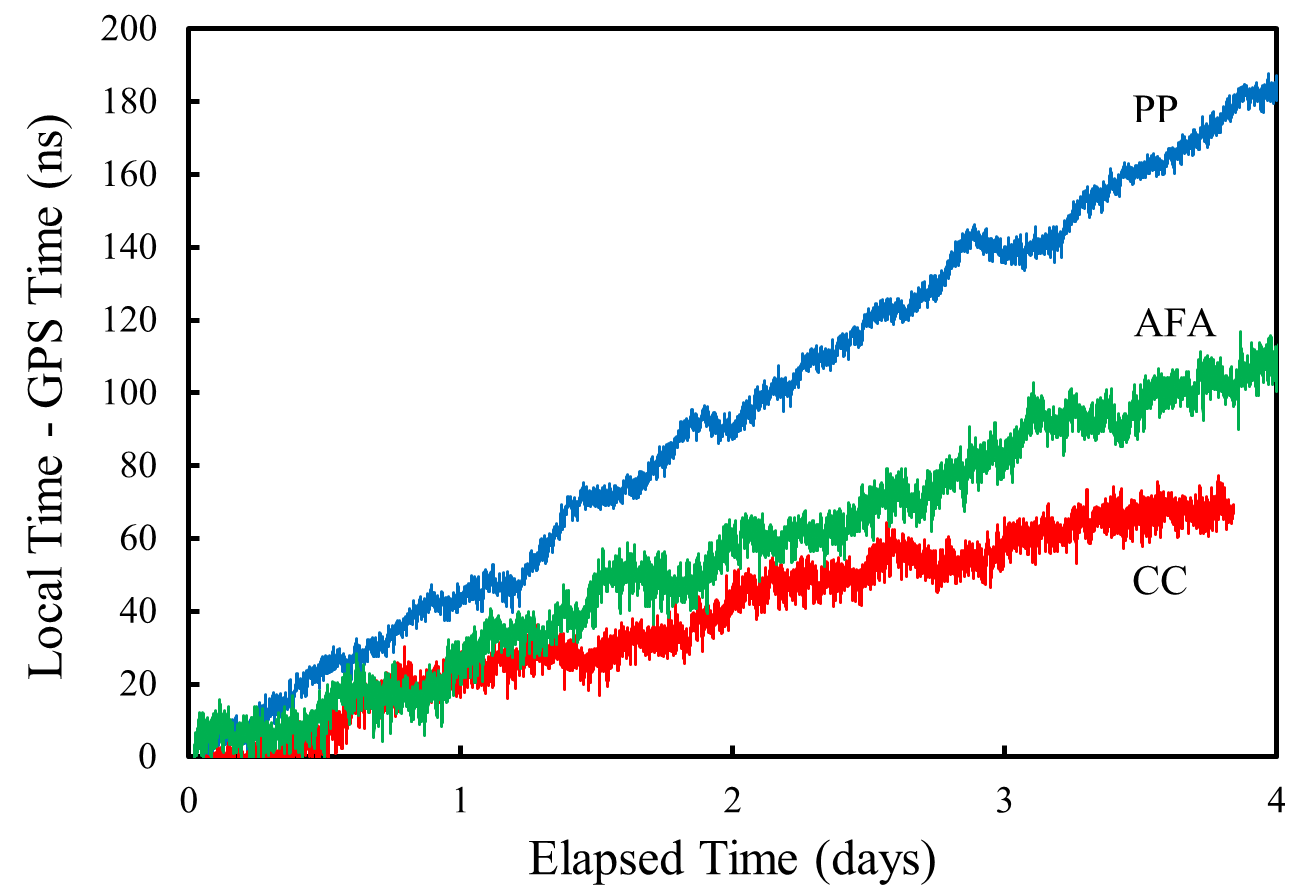}
\caption{\label{fig:dataplot} The difference between local time, as measured by an atomic clock, and sea-level-based GPS time is plotted as a function of elapsed time for three locations: (CC) Colorado College, (AFA) US Air Force Academy, and (PP) the summit of Pikes Peak.}
\end{figure}

Sample data from each of the three locations are shown in Fig.~\ref{fig:dataplot}. The results show qualitative agreement with the predictions of gravitational time dilation. The difference between local time and sea-level-based GPS time increases linearly with elapsed time and accumulates more rapidly for clocks at higher elevations. To make a quantitative comparison, we fit a straight line to each data set, and compared the fitted slope, in units of ns/day, to the values predicted by Eq.~\ref{approxeqn1}. The average results were $22 \pm 3$ ns/day, $26 \pm 2$ ns/day, and $49 \pm 2$ ns/day for the measurements at CC, USAFA, and PP, respectively. The quoted statistical uncertainties were computed using the standard deviation of the mean of multiple measurements at each location. The results are somewhat larger than predicted by gravitational time dilation. We attribute the discrepancies to bias in the cesium frequency standard, which arises from well-documented effects for atomic clocks.\cite{shirley2001} The 5071A data sheet\cite{5017Adatasheet} specifies a long-term stability of less than about $5 \times 10^{-14}$ over a period of five days, or a bias of approximately 4 ns/day. The magnitude of this bias was confirmed by monitoring the time differences between three individual cesium clocks at the same location over a period of about five days. Thus, within the experimental uncertainty, our results are consistent with the predictions of general relativity. Technical details of our experiment, including data acquired at four additional elevations in Colorado, are provided in an in-depth analysis published separately.\cite{burns2017}

\section*{Discussion}

Although the effect of terrestrial time dilation is small, it can be significant for certain applications. One fruitful topic of discussion with students is the implications for satellite operations. All satellites are subject to a time discrepancy between on-board clocks and ground-based systems, but this is an especially critical consideration for GPS satellites. GPS satellites orbit at an altitude of $20,180$ km, where the acceleration due to gravity is only $0.57$ m/s$^2$.  They also orbit at a speed of approximately $3.9$ km/s relative to the surface of the Earth.  By integrating Eq.~\ref{approxeqn1} to GPS altitude, one can show that GPS clocks tick $45.6$ $\mu$s per day faster than clocks on Earth's surface due to gravitational time dilation. Special relativistic effects, on the other hand, cause GPS clocks to tick $5.6$ $\mu$s per day \textit{slower} than clocks on the surface, according to Eq.~\ref{approxeqn2}. This leads to a total time discrepancy of about $40$ $\mu$s per day, which must be accounted for in GPS operation. Further complicating the situation, GPS orbits are not perfectly circular, so the gravitational and special relativistic effects vary over the course of a day. In order for the GPS location algorithms to work correctly, all satellites must agree with each other and with clocks on the surface. If the relativistic effects are not compensated, the GPS system would be useless.\cite{ashby2003}

While the results of our test are not groundbreaking, our aim was to show that advanced modern physics concepts can be tested, and their relevance appreciated, by undergraduates. Air Force Academy cadets and Colorado College students were involved in all phases of this work, including gathering and analyzing data, presenting results at conferences, and assisting in the writing of this paper. Hopefully this short study will also motivate other students to consider careers in Science, Technology, Engineering, and Math (STEM) fields. Dedicated displays of ongoing relativity testing are being established near physics classrooms at both USAFA and Colorado College. The experiment has also been discussed at USAFA in an introductory physics course involving a few hundred cadets. The displays and discussions emphasize the impact of subtle relativistic effects on technological systems such as GPS. Aside from the military and commercial navigation applications, other uses of GPS that could be discussed include: time-stamping financial transactions and synchronizing power-line nodes for fault detection\cite{mcneff2002}; monitoring water distribution of Earth\cite{tapley2004}; and tests of fundamental physics.\cite{cacciapuoti2009} We hope to kindle excitement in students for the beauty, strangeness, and applicability of Einstein's theory of relativity. 

\begin{acknowledgments}
We gratefully acknowledge the support of Brian McGarvey and the RX research group at Fort Meade for providing access to the Cs atomic clocks. Special thanks go to John Bristow for arranging the use of the facilities at the United States Army Pikes Peak Research Laboratory. This project was funded by grants from the Mellon Foundation and CC research funds. J.F.S. acknowledges support from the Air Force Office of Scientific Research and the National Science Foundation (Grant No.\ 1531107).
\end{acknowledgments}

\end{document}